\documentclass[prb,twocolumn,superscriptaddress,floatfix]{revtex4}
\usepackage{graphicx}

\begin{document}

\title{Momentum Resolved tunneling in a Luttinger Liquid}

%%%%%%% author group: %%%%%%

\author{S.~A.  Grigera\thanks{Present address: School of Physics and
Astronomy, University of St. Andrews, St. Andrews KY16 9SS, United
Kingdom.}} 
\affiliation{School of Physics and Astronomy, University of
Birmingham, Birmingham B15 2TT, United Kingdom.}

\author{A.~J.  Schofield}
\affiliation{School of Physics and Astronomy, University of
Birmingham, Birmingham B15 2TT, United Kingdom.}

\author{S. Rabello}
\affiliation{Department of Physics, Rice University, Houston, TX
77251-1892}

\author{Q. Si}
\affiliation{Department of Physics, Rice University, Houston, TX
77251-1892}

%%%%%%%%%%%%%%%%%% Abstract     %%%%%%%%%%%%%%%%%%

\begin{abstract}
We consider momentum resolved tunneling between a Luttinger liquid and
a two dimensional electron gas as a function of transverse magnetic
field.  We include the effects of an anomalous exponent and Zeeman
splitting on both the Luttinger liquid and the two dimensional
electron gas.  We show that there are six dispersing features that
should be observed in magneto-tunneling, in contrast with the four
features that would be seen in a non-interacting one dimensional
electron gas.  The strength of these features varies with the
anomalous exponent, being most pronounced when $\gamma_\rho=0$. We
argue that this measurement provides an important experimental
signature of spin-charge separation.
\end{abstract}

\maketitle

%%%%%%%%%%%%%%%%%% Introduction %%%%%%%%%%%%%%%%%%

\section{Introduction}

Haldane's Luttinger-liquid hypothesis~\cite{haldane_1981a}---that all
one-dimensional metals are adiabatically continuous with the
Tomonaga-Luttinger model~\cite{tomonaga_1950a,luttinger_1963a}---has
underpinned our current understanding of the metallic state in
one-dimension. The low-energy properties of the metal are
characterized by separate spin and charge velocities ($v_\sigma$ and
$v_\rho$ respectively) and, at most, two further anomalous exponents
($\gamma_\sigma$ and $\gamma_\rho$).  The low-energy excitations are
completely different from those of the non-interacting electron gas.
The one-dimensional metal is described in terms of spinons and holons
rather than quasi-electron-like excitations. As a result the low
energy spectrum has no overlap with the corresponding non-interacting
one and the metal is therefore a non-Fermi liquid~\cite{schofield_1999a}.

Although much is know theoretically about the properties of a
Luttinger liquid (see, for example, Voit in
Ref.~\onlinecite{voit_1995a}), experimental verification of these
ideas is on-going.  A wide variety of measurements have been performed
and interpreted within the Luttinger-liquid framework. These include
work on the quasi-1D organics, inorganic charge-density wave
materials, semiconductor quantum wires and edge states in the
fractional quantum Hall
regime~\cite{chang_1996a,chang_2001a}. However, most of these
experiments have focused on identifying the anomalous
exponents. Experiments which directly probe the separation of charge
and spin in one dimension have proved to be more challenging. Arguably
the most convincing measurements have been those of angle-resolved
photoemission in metals~\cite{segovia_1999a} and
insulators~\cite{kim_1996a,fujisawa_1999a}. Nevertheless, there
remains a need for a low-energy probe of the excitation spectrum of
the Luttinger liquid.

In a recent Letter~\cite{altland_1999a}, Altland {\it et al.} proposed
a novel spectroscopy of the Luttinger liquid state using
magneto-tunneling.  They showed how the tunneling conductance between
a quasi-1D metal and a two-dimensional electron gas (the spectrometer)
responds to a transverse magnetic field and allows features associated
with the spinon and holon dispersion to be resolved. This
momentum-conserving tunneling spectroscopy then provides a method of
determining the low energy spectrum and identifying features
associated with separate spin and charge excitations. In that Letter
the authors considered, for simplicity, the special case where the
anomalous exponents $\gamma_\rho$ and $\gamma_\sigma$ were both equal
to~0. They also ignored the effect of the magnetic field on the
spectral functions. The magnetic field was assumed to simply tune the
relative momentum of the tunneling electron as it moves between the
one-dimensional metal and the two dimensional electron gas. This same
tuning can also be achieved by changing the carrier density (and hence
$k_F$) in either the wire or the two-dimensional electron
gas. Experimentally, using a transverse magnetic field is likely to be
by far the easiest way of tuning the intra-chain momentum.  The work
of Altland {\it et al.} raised two further interesting questions that
we will address here. Firstly, how sensitive are the tunneling results
to the value of the anomalous exponent? Perhaps more importantly,
would the Zeeman splitting of Fermi-liquid quasiparticles give rise to
two features and thereby mimic the spin-charge separation that the
experiment was supposed to resolve.

In this paper we revisit the idea of momentum dependent tunneling and
solve for the tunneling conductance for arbitrary anomalous exponent
$\gamma_\rho$. (The other exponent, $\gamma_\sigma$ is equal to 0 in
any rotationally symmetric system.) Recently two of the present
authors~\cite{rabello_2000a} have also calculated the change in the
spectral function Luttinger liquid spectral function due to a magnetic
field. Using this result, we have now computed the tunneling
conductance beyond the restrictions of
Ref.~\onlinecite{altland_1999a}. We find that the signature of
magneto-tunneling into a Luttinger liquid is radically different from
that in a non-interacting one dimensional metal and the magnetic field
reinforces this difference. The dispersion of spinons and holons may
be separately identified from sharp features in the tunneling
conductance. However, these features become less singular as the
anomalous exponent varies away from unity.

The outline of the paper is as follows. We begin by establishing the
formalism for magneto-tunneling when momentum parallel to the wire is
conserved. We then introduce the spectral functions for arbitrary
$\gamma_\rho$ but initially ignore any change induced by the magnetic
field. Next, we show how the magnetic field can straightforwardly be
taken into account within this formalism and we compute the tunneling
conductance. Finally we discuss current attempts to perform this
magneto-tunneling spectroscopy.

\section{Overview}
\begin{figure}
\includegraphics[width=\columnwidth]{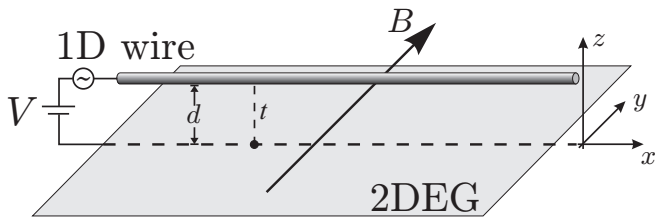}
\caption[fig1]{Schematic representation of the tunneling geometry and
field orientation for the magneto-tunneling experiment.}
\label{schema}
\end{figure}

One of the advantages of this type of spectroscopy of correlated
metals is that the theoretical interpretation of the experiment is
extremely straight-forward. We can therefore sketch our main findings
before detailing more formally how they arise. The tunneling current
between two metals is determined by the rate at which real physical
electrons can hop between them. Since each metal is, in general, a
complex interacting system the physical electron is not an eigenstate
of either metal. Thus the hopping rate is determined not just by the
coupling between the two metals but by the overlap between a physical
electron and the underlying eigenstates of the metallic state. The
probability that a physical electron has a given energy and momentum
in the metallic state is characterized by the spectral
function---which contains all the information about this overlap. In
the geometry considered in this paper, the tunneling electron moves
through a finite voltage and transverse to an applied magnetic
field~(see Fig.~\ref{schema}). Thus the electron changes its energy
(by $eV$) and also its momentum (by $\hbar q_B$ to be defined later)
along the direction of the interface due to the Lorentz force acting
on it during the tunneling process. The crucial assumption that we
make in this paper is that the tunneling barrier is smooth so that
momentum parallel to the 1D wire is conserved (up to this change due
to the Lorentz force).
%%%%%%%%%%%

So, in summary, the tunneling current is determined by the joint
probability that an electron with spin $\sigma$ may be found on one
side of the tunnel barrier with momentum $k_x$ and energy $\omega$ and
that there is an empty electron state with spin $\sigma$ and with
momentum $k_x+q_B$ and energy $\omega + eV$ on the other side of the
barrier.  This is then integrated over all $k_x$ and $\omega$ to get
the final current.  However, what makes this measurement potentially
useful is that in many interacting systems these probabilities (or
spectral functions) contain well defined features that in turn relate
to the true eigenstates of the interacting system.

\begin{figure}
\includegraphics[width=\columnwidth]{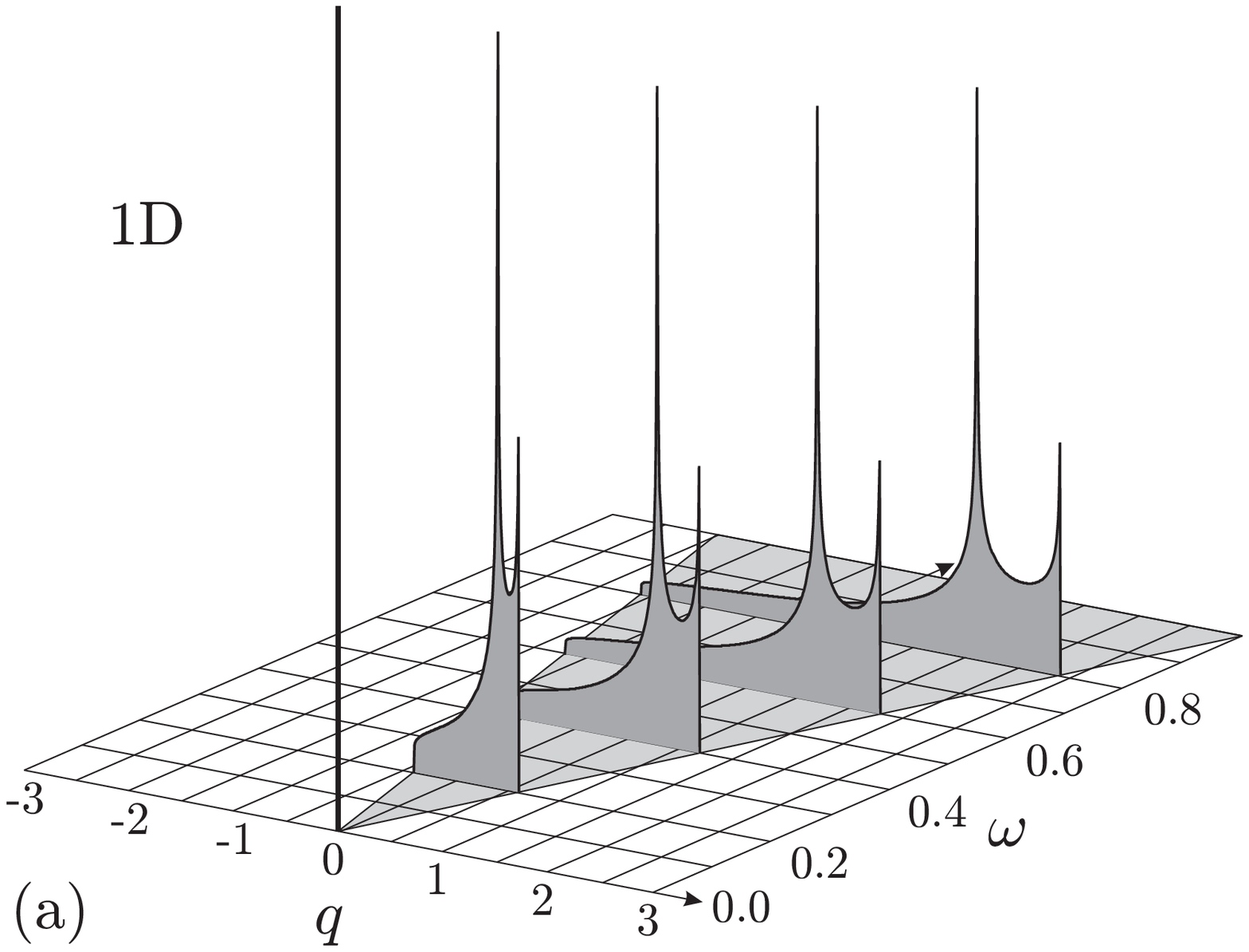}
\includegraphics[width=\columnwidth]{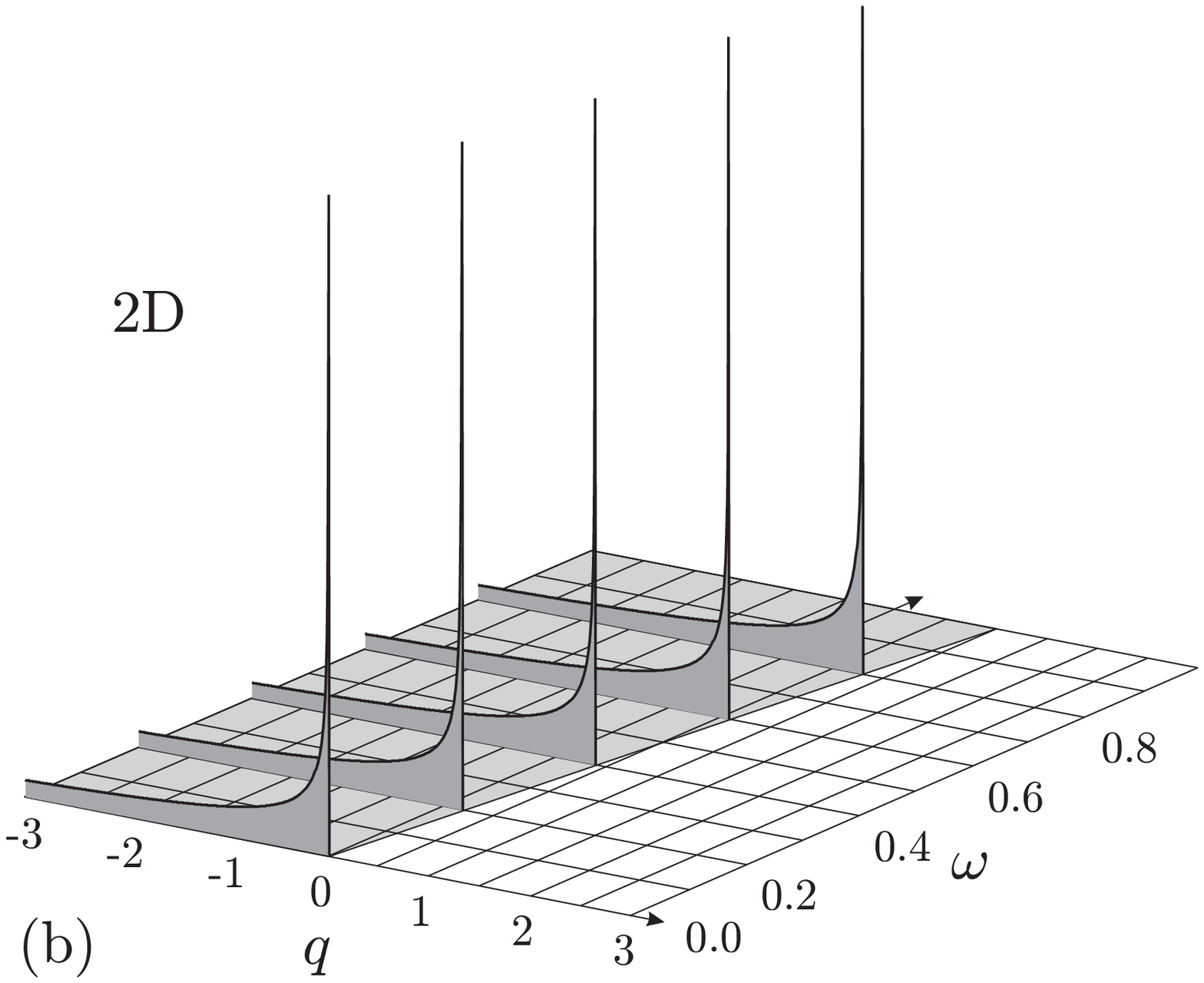}
\caption[fig1]{(a) The electron spectral function in a Luttinger
liquid for $\omega \geq 0$ and $\gamma_\rho=0.05$. The two dispersing
singular features reflect the spinon and holon excitations.  (b) The
electron spectral function for a two dimensional Fermi liquid
integrated over $k_y$ (perpendicular to the 1D wire). The single
singular feature identifies the dispersion of the electron-like
quasiparticle.}
\label{Akw}
\end{figure}

In this paper we consider there to be a Luttinger liquid on one side
of the tunnel barrier. In a Luttinger liquid we know theoretically
that there should be features in the electron spectral function
related to the underlying excitations: spinons and holons. The
dispersion of the spinons and holons can be seen in Fig.~\ref{Akw}(a)
as {\em two} lines of singularities of the spectral function in the
$\omega,q$ plane.  We assume that there is a two-dimensional Fermi
liquid on the other side of the tunnel barrier.  In a Fermi liquid
there are electron-like quasiparticles which are reflected in the
spectral function as a {\em single} line of singularities in the
$\omega,q$ plane [see Fig.~\ref{Akw}(b)].

It is this profound difference in the nature of the excitations of a
Luttinger liquid compared to a Fermi liquid that the
magneto-tunneling experiment is designed to expose. Essentially the
measurement measures the relative dispersion of the singular features
in the 1D and 2D spectral functions as follows. The tunnel current is
given by the integrated product of the two spectral functions of
Fig.~\ref{Akw} with the magnetic field giving a relative offset along
the $q$ direction. Fig.~\ref{integration-areas} shows that this
product (and hence the current) divides into four distinct regions (a)
to (d) depending on this magnetic field dependent offset and the
dispersion of the singular features in the spectral functions. Thus
the tunneling conductance shows three abrupt features separating the
four regions as a function of applied magnetic field. These features
can be seen in Fig.~\ref{didv}.  Finally when one includes the Zeeman
splitting there are separate tunneling processes for up- and down-spin
electrons and the three abrupt features each split in two giving a
total of six features in the conductance. This is shown in
Fig~\ref{g-z}. The dispersion of these features as the magnetic field
and applied voltage is changed also allows us to determine the
relative velocity of the spinon and holon is shown in
Fig.~\ref{contour}. In the absence of spin-charge separation there
would be only one dispersing singularity in the 1D spectral function
leading to four features in the tunneling conductance as a function of
field. This then is the essence of our results and in the rest of the
paper we give a more precise derivation of them.

%%%%%%%%%%%
%
%

%%%%%%%%%%%%%%%%%% Formalism    %%%%%%%%%%%%%%%%%%

\section{Formalism}

We consider single-electron tunneling between a one-dimensional
interacting electron metal, parallel to the $x$ axis, and a
two-dimensional electron gas in the $xy$ plane separated from the 1D
wire by a distance $d$ along the $z$ direction. A potential difference
is applied between the wire and the two-dimensional electron gas and a
magnetic field is applied in the plane of the two dimensional electron
gas but perpendicular to the wire (along the $y$ direction). The
geometry is shown schematically in Fig.~\ref{schema}. The
appropriate formalism was first derived in the context of
superconductivity~\cite{schrieffer_1963a}. The Hamiltonian for the
system may be written as
\begin{equation}
\hat{H} = \hat{H}_{\rm 1D}(B) + \hat{H}_{\rm 2D}(B) + \hat{H}_{\rm T}
\; .
\end{equation}
This differs from the Hamiltonian considered in
Ref.~\onlinecite{altland_1999a} since we have explicitly allowed a
coupling of the magnetic field to the 1D and 2D electron
systems. Since the field lies in the plane of the 2D electron gas
however the coupling will be via the Zeeman interaction. The orbital
part of the interaction with the magnetic field is included in the
tunneling term
\begin{eqnarray}
\hat{H}_T = \int \! \! d\vec{r} dx \; t(x,\vec{r}) \left[e^{-iedB(x+r_x)/2} 
\hat{\Phi}_{\rm 2D,\sigma}^\dagger(\vec{r}) 
\hat{\Psi}_{\rm 1D,\sigma}^{}(x) \right. \nonumber \\
\left. + {\rm H.c.}\right] \; ,
\end{eqnarray}
Here we have included the transverse magnetic field using the
following gauge $\vec{A}=(0, 0, -B x)$ and the usual Peierls coupling.
The crucial assumption behind this method is that the tunneling
amplitude $t(x,\vec{r}$) for an electron to move from position $x$ in
the 1D to position $\vec{r}$ in the 2D gas is translationally
invariant---{\it i.e.} that the tunnel barrier is smooth---and so may
be written as $t(x-x_{\rm 2D},y_{\rm 2D})$. In momentum space then
this coupling may be written as
\begin{equation}
\hat{H}_T = \sum_{\vec{k}} t_{\vec{k}}
\hat{\phi}^\dagger_{k_x-q_B/2,k_y,\sigma} \hat{\psi}_{kx+q_B/2,\sigma}
+ {\rm H.c.} \;  ,
\end{equation} 
where $q_B = eBd+k_f^{2D}-k_F^{1D}$.
Thus we see that momentum parallel to the 1D wire is conserved up to
the change induced by moving the applied magnetic field. The applied
field then tunes the tunneling momentum of the electron.  Here
$t_{\vec{k}}$ is the Fourier transform of the tunneling matrix element.

Given the tunneling Hamiltonian, it is then straightforward to
determine the tunneling current (see, for example,
Mahan~\cite{mahan_1990a}). Note that our starting Hamiltonian neglects
any interactions between the two dimensional electron gas and the
quantum wire---the only coupling is via single electron
tunneling. Thus with this assumption there are no vertex corrections
in the tunneling current and it can be written directly in terms of
the single electron Greens function for the 2D electron gas and the
quantum wire. The form for the current is most intuitively expressed
in terms of the electron spectral functions for the 2D and 1D systems
respectively
\begin{eqnarray}
I(B,V) = \int d\omega \sum_{\vec{k}} t_{\vec{k}}^2 \left[ f(\omega) -
f(\omega-eV) \right] \nonumber \\
A_{1D}(k_x+q_B/2,
\omega, B) A_{2D} (k_x-q_B/2, k_y, \omega-eV, B) 
\label{I}
\end{eqnarray}

For the purposes of this paper we assume that the tunneling is through
nearest contact only so that $t(x-x_{\rm 2D}, y_{\rm 2D} = t_0
\delta(x-x_{\rm 2D}) \delta (y_{\rm 2D})$. It would be straightforward
to relax this assumption by introducing an additional ``aperture
function'' which would need to be included in the momentum
convolution.

%%%%%%%%%%%%%%%%%% Tunnelling  %%%%%%%%%%%%%%%%%%

\section{Tunneling with general $\gamma_\rho$}

We start by specifying the spectral functions used in the calculation
of the current.  For the two-dimensional system the spectral function
$A_{2D,\eta}$ for energies close to the Fermi energy (and integrated
over the momentum component transverse to the wire) is given
by~\cite{altland_1999a} (see Fig. \ref{Akw})
\begin{equation}
   A_{2D}(q,\omega) = \sqrt{2m} \frac{
                                      \Theta(\omega - q v_F)
                                     }
                                     {
                                      \sqrt{\omega - q v_F}
                                     }.                    
                                        \label{A2D}
\end{equation}
The spectral function of a Luttinger liquid with spin
rotation-invariant interactions ($\gamma_\sigma = 0$) is non-trivial
(see Fig. \ref{Akw}), and a closed and tractable analytic expression
is still lacking in the literature. The asymptotic behavior, on the
other hand, is well characterized \cite{voit_1995a}.  At very small
$q$, the function looks very similar to a spinless fermions' function.
As $q$ is increased two peaks become apparent, a reflection of
spin-charge separation, one at $v_\sigma q$ and another at $v_\rho q$,
where $v_\rho$ and $v_\sigma$ are the velocities of charge and spin
density waves, respectively.  The exponent of the singularity at
$v_\sigma q$ is $2\gamma_\rho -\frac{1}{2}$ and the corresponding one
at $v_\rho q$ is $\gamma_\rho -\frac{1}{2}$.  The function terminates
at negative $q$ at $- v_\rho q$ with a non-singular exponent
$\gamma_\rho$.  The parameter $\gamma_\rho$, which characterizes the
interactions between left and right movers is always positive.  The
case of non-interacting left and right branches, which was studied in
reference \onlinecite{altland_1999a}, corresponds to the case
$\gamma_\rho = 0$.  As $\gamma_\rho$ increases, the power law
divergences gradually weaken into cusp singularities, and the spectral
weight, which for $\gamma_\rho = 0$ is confined within $v_\sigma q$
and $v_\rho q$ is gradually transfered by the electronic correlations
toward higher values of $q$ and $\omega$. The spectral function should
also be invariant to the transformations ($p \to -p; q \to -q$) where
$p=\pm$ labels left and right movers, and ($\omega \to -\omega; q \to
-q$). For $\omega >0$, the case of our interest, a function that has
the correct singularities and asymptotic behavior can be written as

\begin{figure}
\includegraphics[width=\columnwidth]{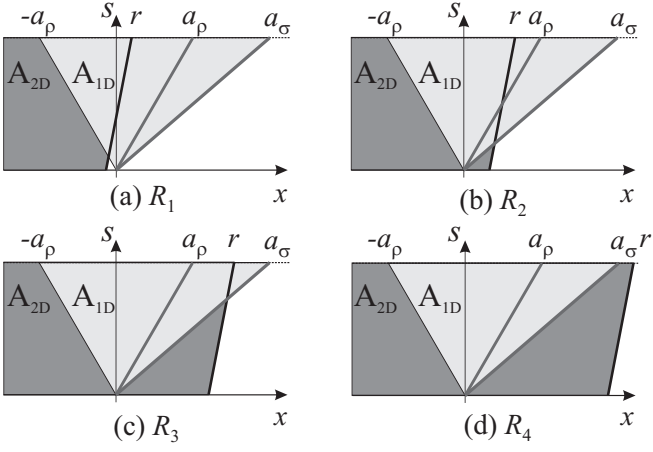}
\caption[fig3]{The tunnel current is determined by the product of the
spectral functions of Fig.~\ref{Akw}(a) and (b) offset in momentum by
$q_B$---an amount proportional to the transverse magnetic field. As
the field is increased the region of overlap goes through four stages
(a) to (d) labeled $R_1$ to $R_4$ in the text.  The light dark and
shaded areas represent, respectively, the areas where the
one-dimensional and the two-dimensional spectral functions have a
finite non-zero value; the thick lines represent the lines of
singularities.}
\label{integration-areas}
\end{figure}

\begin{widetext}
\begin{eqnarray}
A_{1D}(q,\omega) = \frac{W(q,\omega) \Theta(\omega -qv_\sigma)}
{|\omega - v_\sigma q|^{\frac{1}{2}-2\gamma_\rho} |\omega - v_\rho
q|^{\frac{1}{2}-\gamma_\rho}} ~~~{\rm where}~~~ W(q,\omega) =
\cases{\Theta(qv_\rho -\omega)+\Theta(\omega -qv_\rho)c(\gamma_\rho)
&if $\gamma_\rho \ne 0$ \cr \Theta(qv_\rho -\omega) &if $\gamma_\rho =
0$},
\label{A1D}
\end{eqnarray}
for $q>0$ plus a non-divergent term for $q < 0$ and $\gamma_\rho \ne 0$ which is
proportional to
\begin{equation}
\Theta(\omega + v_\rho q)c(\gamma_\rho) (\omega + v_\rho
q)^{\gamma_\rho}. \label{A1Dns}
\end{equation}
This function is not normalizable for $\gamma \ne 0$.  In our case the
consequences of this are merely reduced to an undetermined constant
that multiplies the conductivity for each $\gamma_\rho$.
% This means that it is not possible to
%make quantitative comparison between graphs of conductivities obtained
%for different exponents, but it does not affect any of their
%features.
In this paper we will consider the case
$v_F>v_\rho>v_\sigma$, but it should be noted that the results can be
trivially extended to consider the other possible cases.

Substituting equations \ref{A2D} and \ref{A1D} into \ref{I} we find
for the tunneling current at $T=0$
\begin{equation}
I(V,B) = \frac{4\sqrt{2}I_0(eV)^{\frac{1}{2}+3\gamma_\rho}}{\pi\sqrt{m}v_F}
          \sum_\alpha
          \int_{l_\alpha}^{u_\alpha}  dx
              \int_{L_\alpha}^{U_\alpha}  ds
                   \frac{
                          a_\sigma^{\frac{1}{2}-2\gamma_\rho}  
                          a_\rho^{\frac{1}{2}-\gamma_\rho}
                        }
                        {
                         |sa_\sigma - x|^{\frac{1}{2}-2\gamma_\rho}
                         |sa_\rho - x|^{\frac{1}{2}-\gamma_\rho}
                         | s - x +(r-1)|^{\frac{1}{2}}
                        },
                                                          \label{I2} 
\end{equation}
\end{widetext}
where we have introduced the dimensionless parameters $r = q_B v_F /
eV, a_\rho = v_F / v_\rho$ and $a_\sigma = v_F / v_\sigma$ and the
dimensionless one-dimensional variables $x = q v_F /eV$ for the 1D
wire wave-vector and $s = \omega / eV$ for the frequency. $I_0 = e
|t|^2 m/\pi$ is the natural unit for current in this problem. From
this integral we identify four different regions with different
qualitative behavior, $R_j,~ j= 1 \ldots 4$, corresponding to
different situations of overlap between the one- and two-dimensional
spectral functions. Figure \ref{integration-areas} is an schematic
representation of the relative positioning between $A_{1D}$ and
$A_{2D}$ in the four regions.  In terms of the dimensionless
parameters these are given by $R_1$: $r<1$; $R_2$: $1 \leq r \leq
a_\rho$; $R_3$: $a_\rho \leq r <a_\sigma$; and $R_4$: $r>a_\sigma$.
In each region, and for practical reasons only, the calculation is in
turn split into different integrals of the same integrand. Table
\ref{table-limits} lists the upper and lower limits corresponding to
each different region.  Although the majority of these integrals
cannot be integrated analytically, the asymptotic behavior in the
different regions can be obtained by standard calculus procedures.
The $q<0$ non-singular part of the spectral function for $\gamma \ne
0$ (equation \ref{A1Dns}) only contributes a small featureless onset of
conductivity at $r = -a_\rho$, and a background of finite conductivity
noticeable only for small values of $r$ and big values of
$\gamma_\rho$.

%
%
%%%%%%%%%%%%%

\begin{table}
\caption{Integration limits for equation \ref{I2} in the four
different regions of spectral function overlap shown in
Fig.~\ref{integration-areas}.}
\label{table-limits}
\begin{tabular} {c  c  c  c  c  c}
\hline Region & $\alpha$ & $l_\alpha$ & $u_\alpha$ & $L_\alpha$ &
$U_\alpha$ \\ \hline \\ $R_1$ & 1 & 0 & $r$ & $x-(r-1)$ & $1$ \\ \\ &
1 & 0 & $\frac{r-1}{a_\sigma-1}$ & $sa_\rho$ & $sa_\sigma$ \\ & 2 &
$\frac{r-1}{a_\sigma-1}$ & $\frac{r-1}{a_\rho-1}$ & $sa_\rho$ &
$s+(r-1)$ \\ \raisebox{1.5ex}[0pt]{$R_2$} & 3 & 0 &
$\frac{r-1}{a_\rho-1}$ & 0 & $sa_\rho$ \\ & 4 & $\frac{r-1}{a_\rho-1}$
& 1 & 0 & $s+(r-1)$ \\ \\ & 1 & 0 & $\frac{r-1}{a_\sigma-1}$ &
$sa_\rho$ & $sa_\sigma$ \\ $R_3$ & 2 & $\frac{r-1}{a_\sigma-1}$ & 1 &
$sa_\rho$ & $s+(r-1)$ \\ & 3 & 0 & 1 & 0 & $sa_\rho$ \\ \\ & 1 & 0 & 1
& $sa_\rho$ & $sa_\sigma$ \\ \raisebox{1.5ex}[0pt]{$R_4$} & 2 & 0 & 1
& 0 & $sa_\rho$ \\
\end{tabular}
\end{table}

%
%%%%%%%%%%%%%
\begin{figure}
\includegraphics[width=\columnwidth]{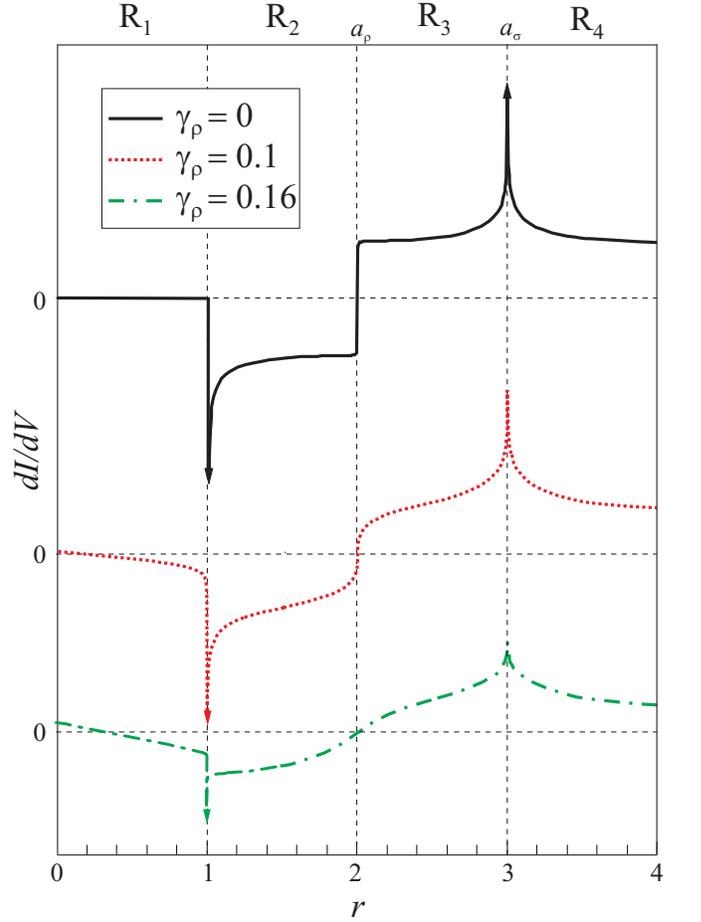}
\caption[fig4]{(color online) Differential tunneling conductance in
the absence of Zeeman splitting shown as a function of dimensionless
magnetic field $r=q_Bv_F/eV$ for different values of the anomalous
exponent, $\gamma_\rho$, and for dimensionless holon velocity $a_\rho
= v_F/v_\rho = 2$ and spinon velocity, $a_\sigma =
v_F/v_\sigma=3$. The graphs have been shifted and re-scaled for
clarity.  The arrows mark divergences. Notice how increasing the
anomalous exponent from the non-interacting value of zero reduces the
three features in the differential conductance.}
\label{didv}
\end{figure}

Figure \ref{didv} shows the differential conductance $G = dI/dV$ as a
function of the dimensionless parameter $r$ at $T=0$. In the following
we discuss the behavior of $G$ in each regime.

$R_1$.--- For $r<1$ and $\gamma_\rho \ne 0$ $A_{2D}$ overlaps with the
non-divergent part of $A_{1D}$, leading to a finite but non-singular
flow of current, with negative differential conductance.  When $r$
reaches the value of 1, the conductance diverges as $g \sim
-(1-r)^{-1/2+3\gamma_\rho}$, where $g = G
\sqrt{E_F}e^{\frac{1}{2}+3\gamma_\rho}V^{-\frac{1}{2}+3\gamma_\rho}/I_0$
is a dimensionless measure of the conductance.  For $\gamma_\rho = 0$
the two areas do not overlap, implying that the current vanishes.

$R_2$.--- For $r>1$, the spectral functions for $\gamma_\rho =0$ start
 overlapping as well, and the conductance diverges as $g \sim
 -(r-1)^{-1/2+3\gamma_\rho}$.  For finite $\gamma_\rho$, however, the cusp is
 also not symmetric (see Fig. 4) since the spectral weight of the
 singularity is different at either side of $qv_\rho$. The behavior of
 the conductance is unaltered up to the boundary to $R_3$, where
\begin{equation}
g(a_\rho^+)-g(a_\rho^-)=\frac{a_\rho}{a_\rho-1}
               \frac{
                 a_\sigma^{1/2-2\gamma_\rho} a_\rho^{1/2-\gamma_\rho}
                     }
                     {
                      (a_\sigma - a_\rho)^{1/2-2\gamma_\rho}
                     }
                      \lim_{r \to a_\rho} (r - a_\rho)^{\gamma_\rho},
\end{equation}
$a_\rho^\pm = a_\rho \pm \delta$, $\delta$ infinitesimal and positive.
This implies that for $\gamma_\rho = 0$, the conductance exhibits a
discontinuity $\Delta$, the magnitude of which is
\begin{equation}
\Delta = \frac{a_\rho}{a_\rho-1}
                          \sqrt{ 
                              \frac{a_\sigma  a_\rho}{a_\sigma - a_\rho} 
                               } \; .
\end{equation}
For every non-zero value of $\gamma_\rho$ this step is rounded off
(see Fig. \ref{didv}) into a continuous function with a pronounced
change at $a_\rho$, the change decreasing progressively as
$\gamma_\rho$ is increased.

$R_3$. --- In this region the differential conductance becomes
positive.  As $r$ approaches the singularity line corresponding to
$a_\sigma$, the boundary with $R_4$, the conductance shows a
pronounced increase.  Again the case of $\gamma_\rho = 0$ is unusual,
since the boundary between $R_3$ and $R_4$ shows a singularity, which
is of logarithmic type:
\begin{equation}
g(r,\gamma_\rho=0) \to_{r\to a_\sigma} -\frac{a_\sigma} {a_\sigma - 1}
\sqrt{ \frac{a_\rho a_\sigma} {a_\sigma - a_\rho} }
\frac{1}{\pi}\ln(a_\sigma - r).
\end{equation}
On the other hand, for any non-zero $\gamma_\rho$, the behavior is
found to be
\begin{eqnarray}
g(r) \to_{r\to a_\sigma} -\frac{a_\sigma} {a_\sigma - 1}
\frac{a_\rho^{1/2-\gamma_\rho} a_\sigma^{1/2-2\gamma_\rho}} {(a_\sigma
- a_\rho)^{1/2-3\gamma_\rho}} \frac{1}{\pi} \nonumber\\ \lim_{r \to
a_\sigma}
{}_2F_1\left[1/2+2\gamma_\rho,1/2+\gamma_\rho,1+2\gamma_\rho,\frac{a_\rho-r}
{a_\rho-a_\sigma}\right],
\end{eqnarray}
which is a non-divergent peak (see Fig. \ref{didv}).

$R_4$. --- The boundary is symmetric in $r$ around $a_\sigma$.  The
asymptotic behavior is $g \sim r^{-1/2}$ for all values of
$\gamma_\rho$.

%%%%%%%%%%%%%%%%%% Zeeman splitting  %%%%%%%%%%%%%%%%%%

\section{Adding a Zeeman splitting}

In this section we consider the effects on the spectral functions of a
Zeeman coupling to the magnetic field.  Very recent work by two of the
authors of this paper\cite{rabello_2000a} has derived the spectral
functions of a Luttinger liquid with a Zeeman term in the Hamiltonian.
  
\begin{widetext}
\begin{equation}
A_{1D}(q,\omega,\varsigma) = \frac{W(q,\omega)\Theta(\omega - q
            v_\sigma -\varsigma B)} { |\omega - v_\sigma q -\varsigma
            B|^{\frac{1}{2}-2\gamma_\rho} |\omega - v_\rho q -
            \varsigma B v_\rho/v_\sigma |^{\frac{1}{2}-\gamma_\rho} }
            \label{A1D-Z}
\end{equation}
where,
\begin{eqnarray} 
W(q,\omega) = \cases{ \Theta(qv_\rho - \varsigma B v_\rho/v_\sigma
-\omega)+\Theta(\omega -qv_\rho - \varsigma B
v_\rho/v_\sigma)c(\gamma_\rho) & if $\gamma_\rho \ne 0$ \cr \Theta(q
v_\rho - \varsigma B v_\rho/v_\sigma - \omega) &if $\gamma_\rho = 0$},
\end{eqnarray}
%\end{widetext}
where $\varsigma$ takes the value of the spin ($\pm 1/2$) times the
Zeeman coupling factor.  We have considered only positive $q$ and
$\omega$, since as we discussed in the previous section the
contribution of $q<0$ is merely to add a finite background of
conductivity. (This background is very small for $\gamma_\rho \sim 0$.)
Furthermore, we have seen that the region of interest, where the
features in the spectral functions are easily distinguished in the
differential conductivity, is restricted to small values of the
anomalous exponent (below $\gamma_\rho \approx 0.2$); for these values
of $\gamma_\rho$ the spectral weight outside the $qv_\rho,qv_\sigma$
region is so small that all non-divergent contributions from $A_{1D}$
outside this interval are negligible.  On the other hand, the effect
of the magnetic field in the two dimensional system is taken into
account by writing the spectral function as
\begin{equation}
A_{2D}(q,\omega,\varsigma') = \frac{\sqrt{m}}{\sqrt{2}} \frac{
      \Theta(\omega - q v_F - \varsigma' B) } { \sqrt{\omega - q v_F -
      \varsigma' B} }, \label{A2D-Z}
\end{equation}
where $\varsigma'$ is the equivalent of $\varsigma$ for the two
dimensional system.

Since we can split the current into the separate contributions of the
two possible values of the spin,
\begin{equation}
I = I_\downarrow+I_\uparrow
\end{equation}
the problem is reduced to the calculation of the integral
%\begin{widetext}
\begin{equation}
I_{\downarrow,\uparrow}(V,B) =
          \frac{4\sqrt{2}I_0(eV)^{\frac{1}{2}+3\gamma_\rho}}{\pi\sqrt{m}v_F}
          \int dq \int d\varepsilon [f(\varepsilon
          -eV)-f(\varepsilon)] A_{1D}(q,\omega,\varsigma)
          A_{2D}(q,\omega,\varsigma').
\end{equation}
\end{widetext}

\begin{figure}
\includegraphics[width=\columnwidth]{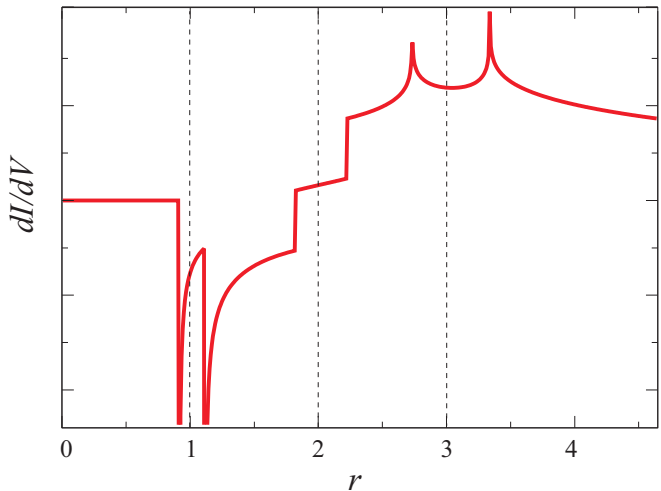}
\caption[fig5]{Differential conductance including the effect of Zeeman
coupling to a magnetic field as well as the orbital effect. The
conductance is shown as a function of dimensionless magnetic field $r$
for the non-interacting exponent, $\gamma_\rho=0$, with dimensionless
holon and spinon velocities, $a_\rho=2$ and $a_\sigma=3$. The Zeeman
coupling ($g$ factor) and the $k_f$ are assumed to be identical in the
1D wire and the 2DEG ($\varsigma=\varsigma'=1$). The features in
Fig.~\ref{didv} appear to spin-split (to different degrees) by the
Zeeman coupling.}
\label{g-z}
\end{figure}

\begin{figure}
\includegraphics[angle =-90,width=\columnwidth]{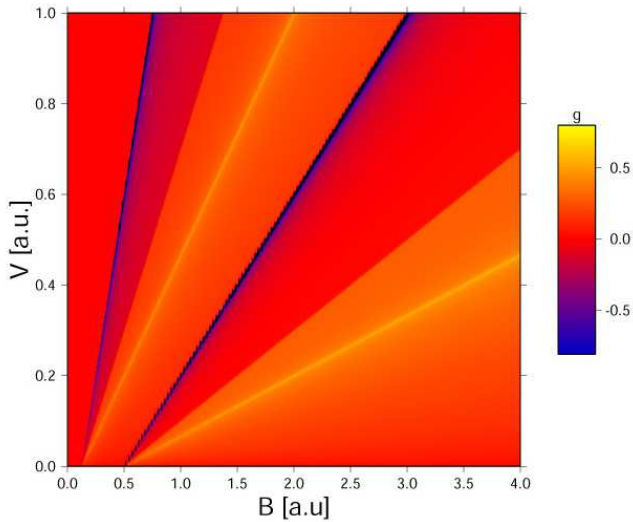}
\caption[fig6]{(color online) Contour plot of the differential
conductance in the presence of a magnetic field as a function of $B$
and $V$ for a small anomalous exponent, $\gamma_\rho=0.05$. Six
dispersing features can be seen---indicative of spin-charge
separation. The curves do not meet at $B=0$ because we do not assume
the magnitude of $k_F$ is the same in the 1D wire and the 2DEG. The
differential conductance will be symmetric under $B \rightarrow -B$ as
tunnelling will then occur via the opposite branch of the Luttinger
and 2DEG spectra.}
\label{contour}
\end{figure}

We can make use of the results of the previous section and simplify
the calculation considerably by defining the spin dependent variable
$q'=q-\varsigma B/ v_s$, and the spin dependent parameter
\begin{equation}
r_{\downarrow, \uparrow} = r \pm \frac{\varsigma'v_\sigma-\varsigma
v_F}{v_\sigma}\frac{B}{eV}.
\end{equation}
The general expressions for the current and the differential
conductance then reduce to
\begin{eqnarray}
I_{\rm Z}=\frac{1}{2} [I(r_\uparrow)+I(r_\downarrow)]\\
g_{\rm Z}=\frac{1}{2} [g(r_\uparrow)+g(r_\downarrow)]
\end{eqnarray}
where $I(r)$ and $g(r)$ are the functions for the current and the
conductance derived in the previous section.  Figure \ref{g-z} shows
the differential conductivity as a function of $r$ for
$\gamma_\rho=0$, $a_\rho=2, a_\sigma=3$, $\varsigma=\varsigma'=1$ and
$k_f^{2D}=k_f^{1D}$.  The different degree of field splitting of the
different features in the conductivity can clearly be seen.

%
%%%%%%%%%%%%
%
%
%%%%%%%%%%%%%
%

%%%%%%%%%%%%%%%%%% Conclusions  %%%%%%%%%%%%%%%%%%

\section{Conclusions}
Having calculated the generalized form of the magneto-tunneling
conductance we see that the key signature of the new types of
excitation in a Luttinger liquid is revealed in the appearance of six
features which disperse with applied field. Loosely this may be viewed
as the allowed transitions between spin-split spinon and holon
excitations and the spin-split electron in the two-dimensional
electron gas. However, this differs dramatically from the case of
electron-like excitations in the one dimensional metal which would
display four features. So this directly addresses the issue as to
whether Zeeman splitting and spin-charge separation are
distinguishable in this experiment---they are.

The effect of an anomalous exponent is more subtle. The original
proposal of Ref.~\onlinecite{altland_1999a} took the case of
$\gamma_\rho = 0$ for calculational simplicity. The more general
treatment given here shows that this is, in fact, the most singular
case and other values for the anomalous exponent leads to less
pronounced effects.  Nevertheless, if $\gamma_\rho$ is not too far
from zero, there will still be six clearly distinguishable features. In
the Luttinger model, $\gamma_\rho$ comes from inter-branch processes,
while spin-charge separation is due primarily to forward scattering
intra-branch effects.  Thus it is possible that spin-charge separation
and an anomalous exponent not far from the non-interacting value of
one could co-exist in real quantum wires.

The role of the anomalous exponent in weakening the tunneling
singularities has implications for other forms of momentum resolved
tunneling experiments. Recently Carpentier {\it
et al.}~\cite{carpentier_2001a} have analyzed the tunneling conductance
between two Luttinger liquids in a magnetic field (though without
including the Zeeman effects as is done here). Again the tunneling
current can be viewed as a convolution but now of two Luttinger
liquid spectral functions. An anomalous exponent which differs from
the non-interacting value will weaken the singularities in both
functions in the convolution and will be doubly detrimental to
features in the conductance. Thus we believe that using a two
dimensional Fermi liquid (2DES) as the spectrometer, as described in
this paper, optimizes the probability of seeing the dispersing
features of the Luttinger liquid. This is because Fermi liquid theory
will always guarantee a square-root singularity in its spectral
function (after integration over the transverse momentum) which is the
best one can do.

The experimental challenges in carrying out this experiment should not
be underestimated. We rely on a number of assumptions. The most
obvious is that tunneling is occurring uniformly along the 1D to 2D
interface rather than via point-like tunneling. The
test for whether an experiment is in this regime comes from the
magnetic field dependence. With point-like tunneling, one would expect
only weak field dependence of the tunneling current since momentum
would no longer be conserved along the wire.  Experiments using MBE
grown interfaces have shown~\cite{kardynal_1997a} that the tunnel
barriers can be sufficiently well controlled to preserve momentum
conservation along the wire during tunneling, hence we believe that
semiconductor fabricated quantum wire to 2D metal interfaces will be
the most promising candidate. 

The second assumption is that the two dimensional system is a
well-controlled Fermi liquid with a large electron weight in the
quasiparticle $Z \sim 1$. This ensures that the overlap between the
electron and the excitations in the 2D spectrometer are large.  Again,
estimates from semiconductor two-dimensional electron gases (2DEGs)
suggest that this is not implausible.

Our final assumption is that the rate limiting step in the experiment
is the tunneling process between the 1D and 2D systems. This requires
a clean quantum wire with no impurities breaking the wire up into
smaller pieces. Early results suggest that this may be causing
problems in trying to implement this experiment in semiconductor
devices~\cite{kardynal_1997a}. There the quantum wire is made by
`pinching off' a channel in a two-dimensional electron gas with an
applied gate voltage. This pushes the one dimensional sub-bands
through the chemical potential until only one remains active. Initial
experimental results reveal magneto-tunneling occurring when multiple
sub-bands are conducting but the wire becomes insulating in the last
sub-band. This is presumably due to impurities blocking conduction (an
interesting process in itself)~\cite{kane_1992a}. Ultimately, we
believe that this should be viewed as a 
challenge rather than a fundamental flaw in the
experiment. However, it also suggests that we should look at
alternative realizations of this experiment. One possibility is to use
carbon nanotubes as quantum wires since these have already been used to
demonstrate Luttinger liquid-like behavior~\cite{bockrath_1999a} via
point tunneling. If a suitable interface could be found with a two
dimensional conventional metal this would be a good alternative
candidate for the magneto-tunneling measurement.

Finally we should point out that, although we have used a magnetic
field for tuning the relative momentum between the wire and
spectrometer, this is not the only method. Using a semiconductor 2DEG
one could back-gate the device and control the carrier concentration,
and hence, $k_F$, in the spectrometer. This gate voltage would then
provide the momentum tuning via the difference in $k_F$ between the
wire and the 2DEG. Such a method could be used in cleaved edge
overgrowth devices where the tunnel barrier to the quantum wire is
in the plane of the 2DEG~\cite{picciotto_2000a}.  Using $k_F$ to tune the
momentum would, of course, mean there is no need to consider Zeeman
coupling. However, the results may be complicated by any carrier
concentration dependence of the 2DEG on its Fermi liquid properties,
or indeed on the parameters of the Luttinger liquid which may be
renormalized via screening from the 2DEG.

To summarize, we have considered momentum-conserving tunneling between
a Luttinger liquid and a two dimensional conventional metal. We have
shown how a transverse magnetic field can be used to tune the relative
momentum of the tunneling electron. This then provides a direct
measure of the spectral function of a Luttinger liquid via its
convolution with that of a conventional Fermi liquid. The signatures
of spin-charge separation are revealed as features in the tunneling
conductance and we have shown they vary as a function of Zeeman
splitting and anomalous exponent. The advantage of this experiment is
that it can be performed with high resolution compared to other probes
of the spectral function such as angle resolved photo emission. Also
the experiment has a very straightforward theoretical interpretation
and hence, if successful, is an unambiguous detector of spin-charge
separation.  We have also discussed the prospects of performing such
an experiment.

%%%%%%%%%%%%%%%%%% Acknowledgements %%%%%%%%%%%%%%

\begin{acknowledgments}

We have benefited from useful discussions with M. W. Long and L. Macks
and we thank A. P. Mackenzie for a critical reading of the manuscript.
We are grateful for the financial support of the Leverhulme Trust, The
Royal Society and NATO Collaborative Research Grant No. 971072. Two of
us (S.R. and Q.S.) have been  supported in part by the Robert A. Welch
Foundation, NSF Grant No. DMR-0090071, and TcSAM.

\end{acknowledgments}

%%%%%%%%%%%%%%%%%% Bibliography %%%%%%%%%%%%%%%%%%

\bibliography{paperdatabase.bib}
\bibliographystyle{apsrev}

\end{document}